\documentclass[traditabstract]{aa}

\usepackage{graphicx}
\usepackage{multirow}
\usepackage{natbib}
\usepackage{adjustbox}
\usepackage{txfonts}
\usepackage{lipsum} % for dummy text only
\usepackage{mathrsfs}
\usepackage{breqn}
\usepackage{linenoaa}

\usepackage[colorlinks = true,linkcolor = blue,urlcolor = blue,citecolor = blue]{hyperref}
\usepackage{xcolor}
\definecolor{yellowgray}{rgb}{0.90, 0.90, 0.2}
\definecolor{bluegray}{rgb}{0.20, 0.60, 0.80}
\definecolor{palered}{rgb}{0.99, 0.40, 0.5}
\definecolor{darkgray}{rgb}{0.35, 0.35, 0.35}
\definecolor{darkgrayb}{rgb}{0.75, 0.75, 0.75}
\definecolor{palegray}{rgb}{0.96, 0.96, 0.96}
\begin{document} 

\title{Infrared spectropolarimetry of a C-class solar flare footpoint plasma. I}
\subtitle{Spectral features and forward modelling}

\author{Vashalomidze Z.\inst{1,2,3} \and Quintero Noda C.\inst{4,5} \and Zaqarashvili, T. V.\inst{6,2,3} \and Benko M.\inst{1} \and Kuridze D.\inst{7,3} \and G\"om\"ory P.\inst{1} \and Ryb\'ak J.\inst{1} \and Lomineishvili S.\inst{1,2,3} \and Collados M.\inst{4,5} \and Denker C.\inst{8} \and Verma M.\inst{8}  \and Kuckein C.\inst{4,5} \and Asensio Ramos A.\inst{4,5}
}

\institute{Astronomical Institute, Slovak Academy of Sciences, 05960 Tatransk\'a Lomnica, Slovak Republic\\
            \email{zurab@astro.sk; zurab.vashalomidze.1@iliauni.edu.ge}
            \and Evgeni Kharadze Georgian National Astrophysical Observatory, Mount Kanobili, 0301 Abastumani, Georgia\
            \and Space Research Centre, Ilia State University, Kakutsa Cholokashvili Ave 3/5, 0162 Tbilisi, Georgia\
            \and Instituto de Astrof\'\i sica de Canarias (IAC), E-38205 La Laguna, Tenerife, Spain\
            \and Departamento de Astrof\'\i sica, Universidad de La Laguna, E-38206 La Laguna, Tenerife, Spain\                       
            \and AG, Institute of Physics, University of Graz, Universit\"atsplatz 5, 8010 Graz, Austria\
            \and National Solar Observatory, 3665 Discovery Drive, Boulder, CO 80303, USA\
            \and Leibniz Institute for Astrophysics Potsdam (AIP), An der Sternwarte 16, 14482 Potsdam, Germany\
            }
\date{Received   ; accepted }

\abstract
{
%Traditional abstract here
%
We performed high-spatial resolution spectropolarimetric observations of active region NOAA 13363 during a C-class flare with the Gregor Infrared Spectrograph (GRIS) on 16 July 2023. We examine the coupling between the photosphere and the chromosphere, studying the polarimetric signals during a period that encompasses the decaying phase of a C-class flare and the appearance of a new C-class flare at the same location. We focus on the analysis of various spectral lines. In particular, we study the \ion{Si}{i} 10827~\AA, \ion{Ca}{i} 10833.4 \AA, \ion{Na}{i} 10834.9 \AA, and \ion{Ca}{i} 10838.9~\AA\ photospheric lines, as well as the \ion{He}{i} 10830 \AA\ triplet. GRIS data revealed the presence of flare-related red- and blueshifted spectral line components, reaching Doppler velocities up to $\sim$90 km s $^{-1}$, and complex \ion{Si}{i} profiles where the \ion{He}{i} spectral line contribution is blueshifted. In contrast, the photospheric \ion{Ca}{i} and \ion{Na}{i} transitions remained unchanged, indicating that the flare did not modify the physical conditions of the lower photosphere. We combined that information with simultaneous imaging in the \ion{Ca}{ii}~H line and TiO band with the improved High-resolution Fast Imager (HiFI+), finding that the flare emission did not affect the inverse granulation or nearby plage, in agreement with the results from GRIS. We also complement the previous studies with a forward modelling computation concluding that the \ion{He}{i} spectral line emission reflects a complex response of the flaring chromosphere. Radiative excitation from coronal EUV irradiation, energy deposition by flare-accelerated electrons, and dynamic field-aligned plasma flows likely act together to produce the observed supersonic downflows and upflows. We plan to expand these findings through inversions of the \ion{He}{i} 10830 \AA\ triplet signals in the future.}


\authorrunning{Vashalomidze et al.}

\titlerunning{Infrared spectropolarimetry of a C-class solar flare footpoint plasma}

\maketitle

\nolinenumbers

\section{Introduction} \label{sec:intro}

Among the many events that take place in the solar system, solar flares are one of the most powerful in terms of energy release. These events are described as sudden energy releases that take place within active regions \citep[see, for instance, ][as general references]{Svestka1976,2002tsai.book.....S,2011SSRv..159...19F}. In a very short time, typically within a few minutes, solar flares are able to emit vast amounts of radiation across the entire electromagnetic spectrum, ranging from long radio wavelengths to high-energy gamma-rays. The sudden surge of energy that is released after the flare event is sometimes carried along magnetic loops to the lower solar atmosphere either by accelerated particles, thermal conduction, or magnetohydrodynamic waves \citep[e.g., ][]{Hirayama1974, Aschwanden2005}. Moreover, part of the plasma can rise again, resulting in a plasma that can be observed with downward and upward-moving velocities as time passes \citep{Neupert1968, Fisher1985, Graham_2015}.

The amount of radiation depends on the flare phase, which can be roughly divided into two main phases: the impulsive and the gradual phases \citep[for example, ][]{Svestka1976}. The former phase is the most active part of a flare, where high-energy radiation in the form of hard X-rays, gamma-rays, microwaves, or white light is produced. These types of radiation point towards the fact that electrons and ions are accelerated to high speeds, where they can reach velocities of tens of thousands of km~s$^{-1}$ \citep[see][for more information]{HudsonRyan1995, Zharkova2011, Reames2013}. 

During the impulsive phase, we believe the particles carry the high energy with them towards the lower layers of the solar atmosphere. There, these particles interact with the solar plasma, heating it and making it glow brightly in different wavelengths \citep{2011SSRv..159...19F}. The locations where the mentioned high-energy particles interact with the solar surface are commonly referred to as the footpoints or ribbons of the solar flare. These footpoints mark the ends of the coronal magnetic loops, where most of the energy of the solar flare is transformed into radiation at the chromosphere and photosphere \citep[e.g.,][]{Hirayama1974}. The gradual phase, on the other hand, is a more gentle phase involving a slower but long-lasting release of energy that can be present for much longer, from minutes to even hours, and it is primarily observed as less energetic soft X-rays and ultraviolet emission. Recent observations of flare acceleration sites have employed combined imaging and spectroscopic diagnostics. For example,  \cite{Li2024} used ASO-S \citep{Li_2019} and CHASE \citep{Li_2022} observations to examine white-light flare evolution.  \cite{Chen2024} identified magnetic-bottle structures at loop-top regions where energetic electrons are accelerated. Spectroscopic studies by  \cite{Joshi2025} revealed turbulent flows and non-Gaussian line profiles associated with filament reconnection during two-ribbon flares.

From previous studies, it seems that in order to better understand the mechanisms of energy transfer between the different atmospheric layers during a solar flare, specifically during its impulsive phase, we need to study in detail the lower part of the solar atmosphere. In particular, we need to infer the spatial and vertical distribution of the magnetic field vector as well as the thermodynamics of the different atmospheric layers. This is a task that is only attainable through observations of spectral lines sensitive to different layers of the solar atmosphere. This requirement leads to the need for multi-channel spectropolarimetry, in other words, the simultaneous observation of several spectral lines either in the same sensor or, in the best case, with multiple wavelength regions recorded at the same time inside one or multiple instruments. This approach will provide invaluable data sets for diagnosing the thermal and magnetic properties of the solar atmosphere and, for this reason, the next generation of ground-based solar telescopes, Daniel K. Inouye Solar Telescope \citep[DIKST,][]{Rimmele2020} and the European Solar Telescope \citep[EST,][]{QuinteroNoda2022EST}, have been designed since their early phases to allow and to be optimised to perform these multi-channel spectropolarimetric observations. 

In our case, we aimed to make use of the Gregor Infrared Spectrograph \citep[GRIS,][]{collados2012} installed at the Gregor telescope \citep{schmidt2012a,Kleint2020} to analyse several of the spectral lines that fall in the spectral range centred at 10830~\AA. In many simple slab--model calculations (optically thin), the ratio of the blue component (\ion{He}{i}~10829.081~\AA) to the red (blended) components (\ion{He}{i}~10830.250~\AA\ + 10830.341~\AA) is often taken to be around $1:8$  (around $\pm 0.12 $) \citep{Fisher1964, Akimov2014, Vicente2023}., along with other diagnostic spectral lines, such as the \ion{Si}{i}~10827~\AA\ transition, can provide valuable information on the thermodynamics of the solar chromosphere when affected by the flare activity \citep[e.g.,][]{Sasso2011, Xu2012, Kuckein2015, Kuckein2025}. Previous observations discovered that the intensity profiles of the \ion{He}{i} triplet were enhanced up to emission levels (above the local continuum) due to the flare event \citep{petrie2010, Li2007, Sasso2011, Kuckein2015}. This enhanced emission seems to be associated with the bombardment of non-thermal electrons that reach the chromosphere after the flare process has started \citep{nagai1984, tei2018, tetsu2018}. At the same time, previous observations also show that the \ion{Si}{i} line, mostly sensitive to physical parameters of the lower and middle photosphere, usually experiences less dramatic changes in its intensity profile, both in its shape or the depth of the spectral line \citep{tetsu2018}. This is in contrast with some works describing radiative-hydrodynamic flare models that predict that the \ion{Si}{i} line core intensity can increase appreciably when the heating of the lower atmosphere is included in the simulation scheme \citep{Ricchiazzi1983ApJ272, Ding2005AandA432, Allred2015ApJ809, Reep2016ApJ827}. 

This study presents full Stokes observations of dynamic changes that occurred in both the photosphere and chromosphere of an active region during and after eruptive events associated with two C-class flares. It represents the first step of a series of investigations where we aim to understand the physical mechanisms that trigger flares and how the solar atmosphere responds to such extraordinary phenomena. In this regard, we focus on analysing the spectral features of the spectral lines observed with GRIS, with particular emphasis on studying the response of the \ion{He}{i}~10830~\AA\ triplet to the flaring phenomena and the impact of such phenomena in the lower atmosphere.

\section{Observations and data preparation} \label{sec:observ}

On 2023 July~16, a C4.7 flare (FL1) and a C3.5 flare (FL2) were observed near the active region NOAA~13363 with the instruments Gregor Infrared Spectrograph \citep[GRIS,][]{collados2012} and the improved High-resolution Fast Imager \citep[HiFI+,][]{Denker2023, denker_2018, Denker2018} installed at the 1.5-metre Gregor solar telescope \citep{schmidt2012a, Kleint2020}. The active region was located at heliocentric-Cartesian coordinates ($+728$\arcsec, $-372$\arcsec), corresponding to a heliocentric angle of $55^{\circ}$ ($\mu = 0.57$). 

\begin{figure}[t]
	\centering
	\includegraphics[width=0.5\textwidth]{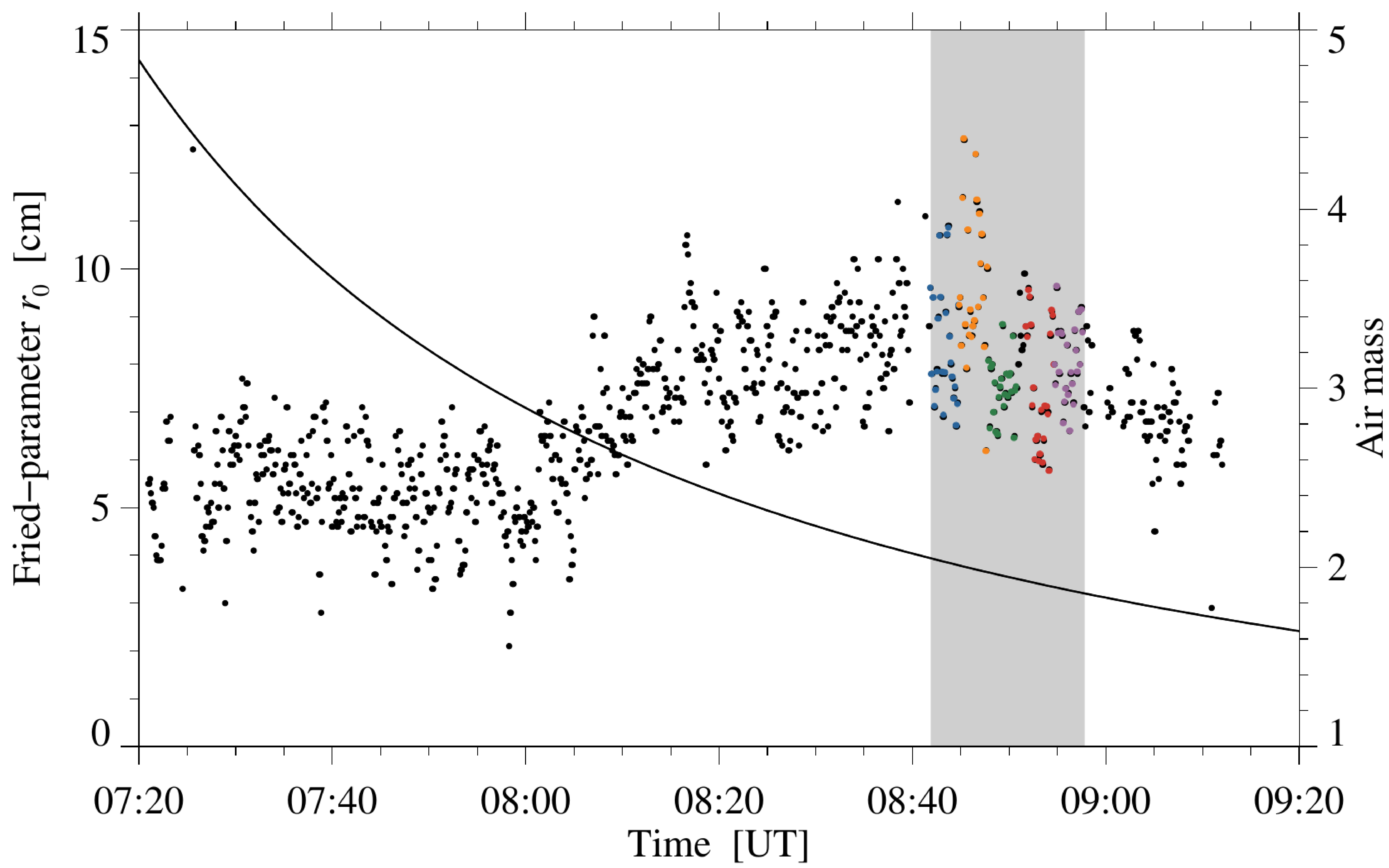}
	\caption{Temporal evolution of the Fried-parameter $r_0$ (black bullets) and air mass (solid line) during the 2023 July16 observations with the 1.5-metre Gregor solar telescope. The light grey rectangle corresponds to the time interval of the HiFI+ observations. The black bullets were taken from the Gregor status monitor and the colour-coded bullets are recorded in the headers of the GRIS scans. The colour-coded bullets refer to the Fried-parameter $r_0$ for each step during the five GRIS scans. The air mass was obtained by using the web interface to the JPL Horizons On$-$Line Ephemeris System.}
	\label{SEEING}
\end{figure}

\begin{figure*}
	\centering
	\includegraphics[width=\textwidth]{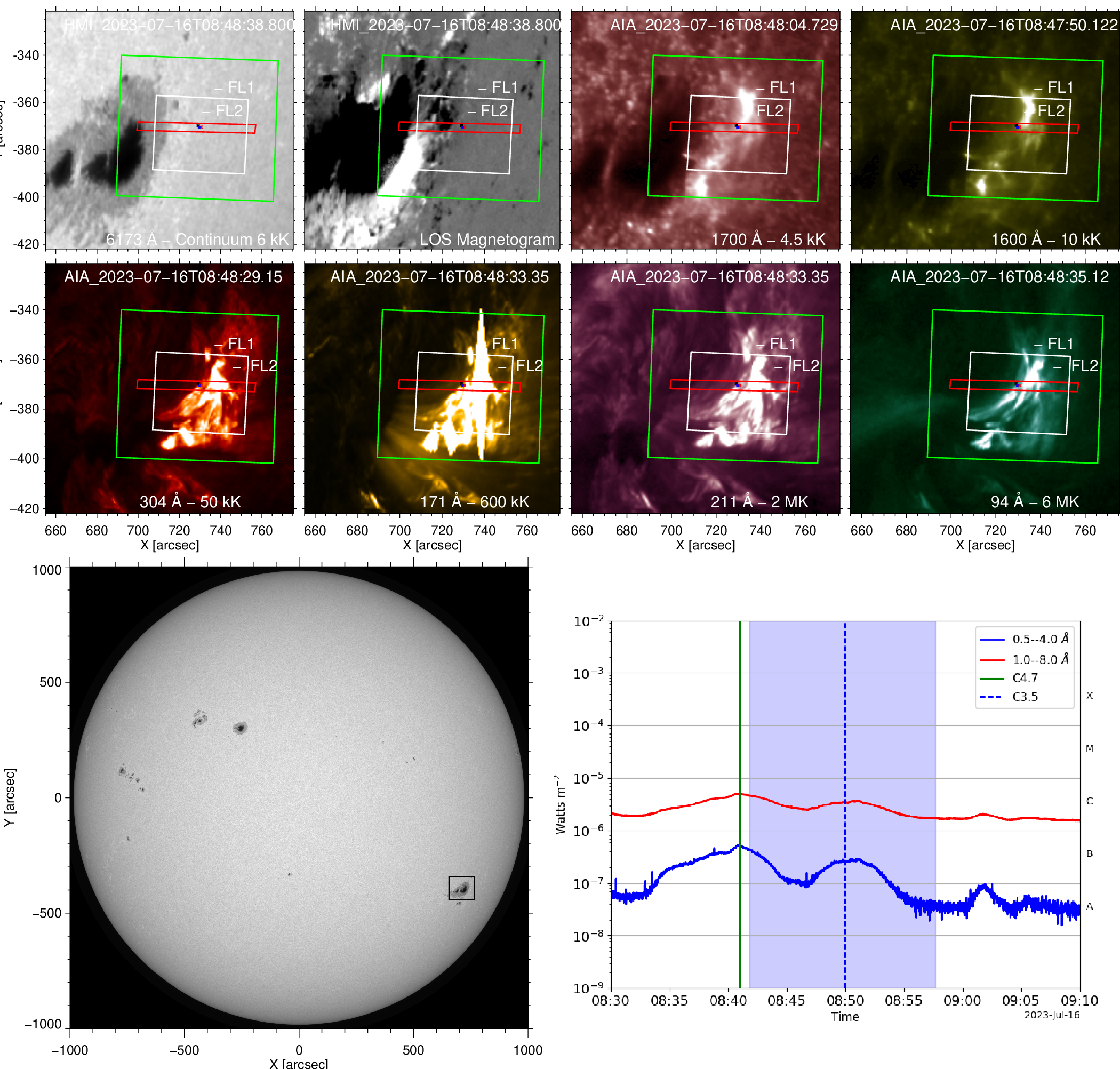}
	\caption{Collection of SDO observations taken at around 08:48~UT. Co-aligned Gregor/GRIS slit-scan and HiFI+ FOVs are shown as red and green (TiO) as well as white (\ion{Ca}{ii}~H) rectangles, respectively. The locations of the first flare (FL1) and second flare (FL2) are labelled in each panel. Bottom left image displays an HMI continuum image of the solar disk at the time of the observation, highlighting with a black square the location of the active region of interest. The lower right panel shows the temporal evolution of the X-ray flux recorded by GOES in the 1.0\,--\,8.0\,\AA\ (red) and 0.5\,--\,4.0\,\AA\ (blue) channels for the period from 08:30~UT\,--\,09:10~UT on July 16, 2023. Gregor's observations are indicated by the blue shaded region. The X-ray maxima are indicated by a solid green line for FL1 and a dotted blue line for FL2, corresponding to the flare peaks.}
	\label{fig:sdo_hmi}
\end{figure*}

Real-time image correction was provided by the Gregor Adaptive Optics System \citep[GAOS,][]{berkefeld2012, berkefeld2010}, which was operated effectively throughout the observing period under favourable conditions. Figure~\ref{SEEING} shows the Fried-parameter $r_0$, measured only when the AO system is locked\footnote{\href{https://status.tt.iac.es}{status.tt.iac.es}}, during a two-hour observing window. The 20-minute time interval of the HiFI+ and GRIS observations is indicated by the light grey rectangle, which is characterised by the good seeing conditions ($r_0 = (6.75 \pm 1.66)$~cm and $r_{0,\,\mathrm{max}} = 12.7$~cm) and a low air mass ($< 2$). The air mass was obtained by using the web interface to the JPL Horizons system.\footnote{\href{https://ssd.jpl.nasa.gov/horizons}{ssd.jpl.nasa.gov/horizons}}

\subsection{Gregor Infrared Spectrograph}

The slit orientation was perpendicular to the nearby solar limb (the slit scan orientation is shown in Figure~\ref{fig:sdo_hmi}). Eight accumulations of four independent modulation states were acquired at each spatial step, with an exposure time of 100 ms per individual image. A total of five scans of 35 steps each were recorded with a cadence of about three minutes per scan. The step size and pixel size along the slit were both 0.135\arcsec, resulting in a field of view (FOV) of $59.4\arcsec \times 4.7\arcsec$. The spectral sampling was 18~m\AA\ pixel$^{-1}$. The observed spectral region encompassed the photospheric \ion{Ca}{i}~10839~\AA\ (deeper photosphere) and \ion{Si}{i}~10827~\AA\ (upper photosphere) lines, which are Zeeman triplets characterised by a Landé factor of $g_{\rm eff} = 1.5$ \citep{balthasar2016}. In addition, the spectral region included the \ion{He}{i} triplet that provides information on the upper chromosphere.
 
The first three scans covered two consecutive flares. The first GRIS scan was obtained from 08:41:50~UT to 08:44:49~UT, corresponding to the post-flare phase of the flare FL1. The second scan was taken from 08:44:49~UT to 08:47:49~UT during the quiescent phase, i.e., the relatively quiet period of lower activity between the more intense flare events. The third scan was obtained from 08:47:49~UT to 08:50:48~UT, corresponding to the initial phase of the flare FL2. In addition, two more scans of the same area were taken from 08:51:41~UT to 08:57:41~UT. However, this work focuses only on the first three scans, which contain the mixture of the decaying phase from the first C-class flare and the impulsive phase of the second C-class flare.

The performed data processing includes photometric corrections, dark current subtraction, flat fielding, polarimetric calibration, and cross-talk removal according to the standard GRIS data reduction software, as described by \citet{collados1999, collados2003}. This process reduces instrumental polarisation to values below about $10^{-3}\,I_c$. The Stokes profiles were then normalised to the local average continuum intensity calculated from quiet-Sun regions within the same field of view, carefully avoiding strong magnetic concentrations. For wavelength calibration, we relied on the telluric lines present in our observed spectral range, which serve as stable absolute references. The calibration procedure followed the standard GRIS data reduction pipeline corresponding to a velocity uncertainty lower than $\sim0.5$~km~s$^{-1}$. Additionally, we adjusted the continuum shape by fitting a polynomial to the ratio of the observed disk-centre gradient and that of the IAG solar atlas spectrum \citep{Reiners2016}, ensuring consistency with high-resolution solar reference data. It is noteworthy that small residual shifts, such as a slight blueshift observed in the \ion{Si}{i} line, are consistent with typical convective blueshifts in the upper photosphere and the limits of wavelength calibration accuracy. The Cassda GUI\footnote{\href{https://gitlab.leibniz-kis.de/sdc/gris/cassda_gui}{gitlab.leibniz-kis.de/sdc/gris/cassda\_gui}} software was used to remove the spikes in the spectral dimension of the Stokes parameters, using the interpolation method. 

Finally, the observations can be downloaded from the GRIS data archive, as per the open data policy.\footnote{\href{https://sdc.leibniz-kis.de}{sdc.leibniz-kis.de}}

\subsection{Improved High-resolution Fast Imager}

%The diffraction limited resolution of Gregor at these wavelengths is 0.055\arcsec\ and 0.097\arcsec, respectively. 

High$-$cadence (about 12~s) and high$-$spatial resolution (plate scale of 0.025\arcsec~pixel$^{-1}$ and 0.05\arcsec~pixel$^{-1}$) observations were obtained with HiFI+ in the \ion{Ca}{ii}~H and TiO channels at 3968~\AA\ and at 7058~\AA, respectively. In HiFI+, the \ion{Ca}{ii}~H filter has a FWHM of 10.8~\AA, and the TiO filter has a FWHM of 9.46~\AA. The filter characteristics are given in Table~1 and Figure~9 of \citet{Denker2023}. The \ion{Ca}{ii}~H filter integrates over both the line core and line wings, leading to a significant contribution from the upper photosphere and lower chromosphere. The line-wing intensity is about half the continuum intensity at the points where the transmission of the filter is 50\%. This mixing reduces the spectral purity of this chromospheric signal. However, small-scale magnetic fields and flare emission still produce a significant contribution to the integrated \ion{Ca}{ii}~H intensity. The TiO bandhead has a strong response to magnetic fields which is absent in the field-free quiet Sun. Thus, it is commonly used for proxy magnetometry in the photosphere, where it benefits from better seeing conditions at longer wavelengths as compared to the Fraunhofer G-band.

In total, 80 datasets captured the leading main spot and the flare from 08:41:53~UT to 08:57:49~UT. Each dataset initially consisted of 500 short-exposure (9~ms) images, of which the best 100 were selected for image restoration using the Kiepenheuer Institute Speckle Imaging Package \citep[KISIP,][]{Woeger2008}. Data calibration and interfaces to image restoration methods are implemented in the sTools data processing pipeline \citep{stools}. The restored images have a size of $1936 \times 1216$ pixels and $1536 \times 1208$ pixels, corresponding to a FOV of $48.2\arcsec \times 30.8\arcsec$ and $76.5\arcsec \times 60.5\arcsec$, respectively. HiFI+ data is available.\footnote{\href{https://gregor.aip.de}{gregor.aip.de}}

\subsection{SDO/AIA+HMI instruments}

We also use magnetogram data from the Helioseismic and Magnetic Imager \citep[HMI,][]{Scherrer2012} and imaging data from the Atmospheric Imaging Assembly \citep[AIA,][]{Lemen2012} instrument onboard the Solar Dynamics Observatory \citep[SDO,][]{Pesnell2012} to trace the impact of the two C-class flares in plasma of higher temperature (see Figure~\ref{fig:sdo_hmi}). We overplot on the SDO images the co-aligned Gregor/GRIS slit-scan (red rectangle) and HiFI+ \ion{Ca}{ii}~H and TiO FOVs (white and green boxes, respectively). 

\begin{figure}
	\centering
	\includegraphics[width=0.5\textwidth]{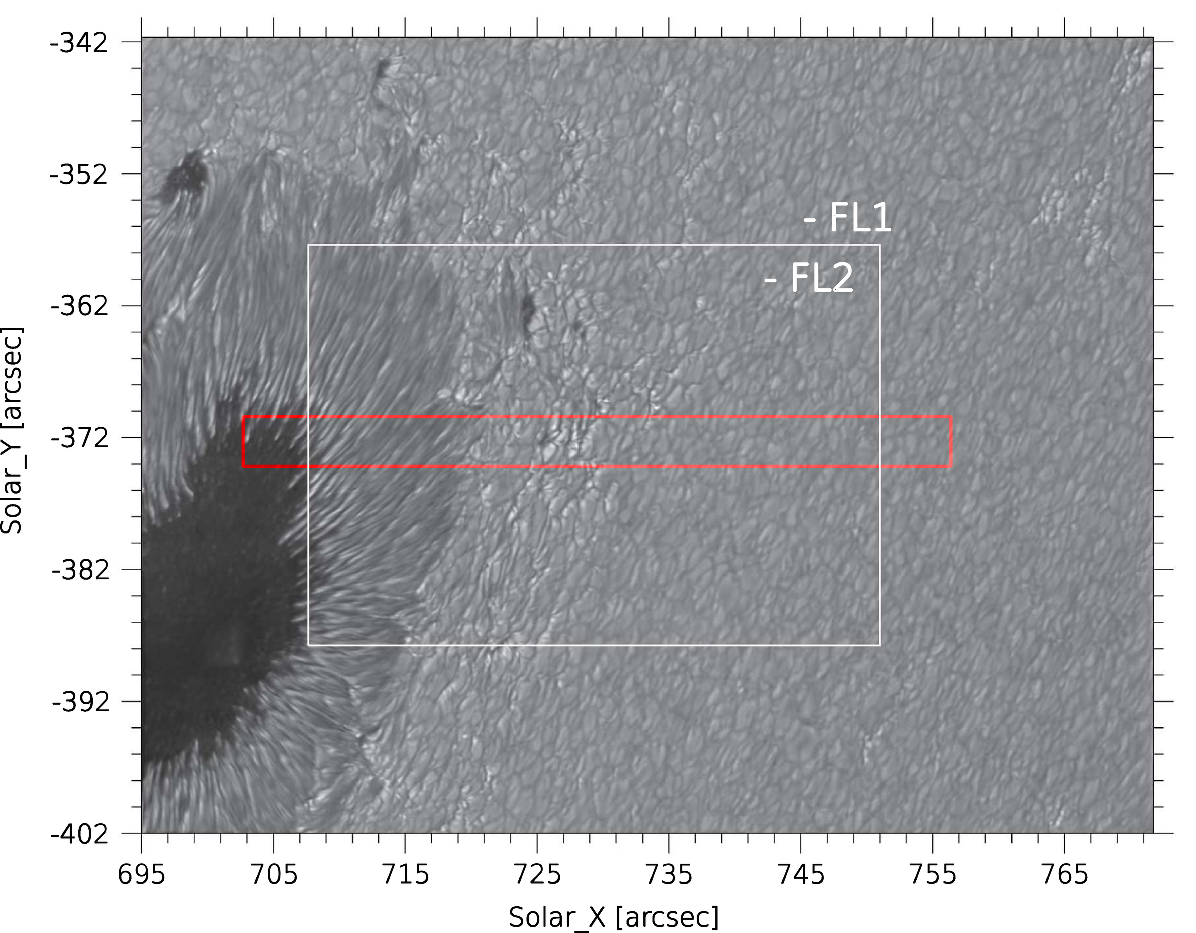}
	\caption{Speckle-restored TiO image taken at 08:45:18~UT at the start of the second flare FL2. The white rectangle indicates the size and location of the speckle-restored \ion{Ca}{ii}~H image shown in Figure~\ref{CAIIH}. The red solid rectangle shows the area covered by the Gregor/GRIS slit-scan.}
	\label{fig:TIO}
\end{figure}

\begin{figure*}
	\centering
	\includegraphics[width=\textwidth]{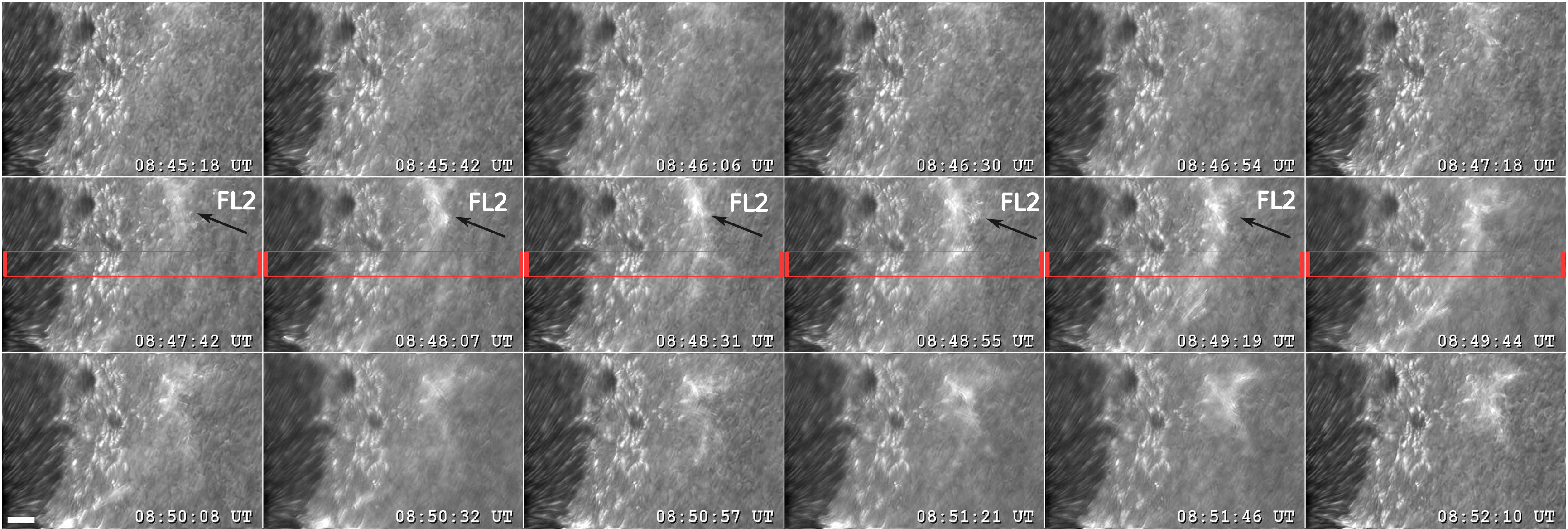}
	\caption{Time series of speckle-restored \ion{Ca}{ii}~H images showing the temporal evolution of the second flare FL2 at a cadence of about 12~s (time interval $-$ 24~s). The small white rectangle in the lower left corner has a size of $5\arcsec \times 1\arcsec$. Gregor/GRIS slit scans are indicated by the red solid rectangle. The black arrow and the label ``FL2'' indicate the identified location of the second flare. Note that the field of view of these HiFI+ \ion{Ca}{ii}~H observations does not cover the location of FL1, which lies outside these panels.}
	\label{CAIIH}
\end{figure*}

We also include the temporal evolution of the X-ray flux in the Geostationary Operational Environmental Satellites (GOES) channels (1.0\,--\,8.0\,\AA\ and 0.5\,--\,4.0\,\AA) in Figure~\ref{fig:sdo_hmi}.  The observation period of Gregor is marked by the blue shaded area, and the maximum intensities of the first flare (FL1) and second flare (FL2) are indicated by solid green and blue dotted vertical lines, respectively. The first spatial scan (08:41:50\,--\,08:44:49~UT) covered the post-flare phase of FL1, while the second flare (FL2) commenced at 08:46:00~UT, peaked around 08:51:00~UT, and ended at 08:56:00~UT. The third scan (08:47:49\,--\,08:50:48~UT) coincided with the rise and maximum phase of FL2. The set of images reveals that the Gregor observations were concentrated near the loop footpoints, where dynamic processes associated with the solar flares are typically more prominent.

\section{Data analysis} \label{sec:results}

After examining the time series of speckle-restored TiO images, we found that they do not show any flare-related intensity variations. Thus, we selected the snapshot with the best image quality to study the spatial distribution of intensity signals in the lower photosphere and plotted it in Figure~\ref{fig:TIO}. We can see that the active region dominates the left side of the FOV, while an area of bright points appears on the right side of the sunspot (around the middle of the FOV). There are also additional small-scale active pores, for instance, at around position $(720\arcsec,\,-362\arcsec)$. However, the occurrence of the flare activity happens further to the right side of the FOV (see labels FL1 and FL2), so it is not straightforward to connect the magnetic activity on the left side with the occurrence of the flares far to the right side.

In the case of the spatial distribution of \ion{Ca}{ii}~H chromospheric intensity signals, we have significant variations with time (see Figure~\ref{CAIIH}). The sunspot appears as a dark feature in comparison with its surroundings, except in the penumbra, where bright filaments can be detected. Close to it, towards the right side of the FOV, we have significant chromospheric activity correlated with the bright points detected in the photosphere. Further to the right, we can see the gradual appearance of the second C-class flare and how the GRIS scan captured part of it (see red rectangle). 

\begin{figure}
	\centering
	\includegraphics[width=0.5\textwidth]{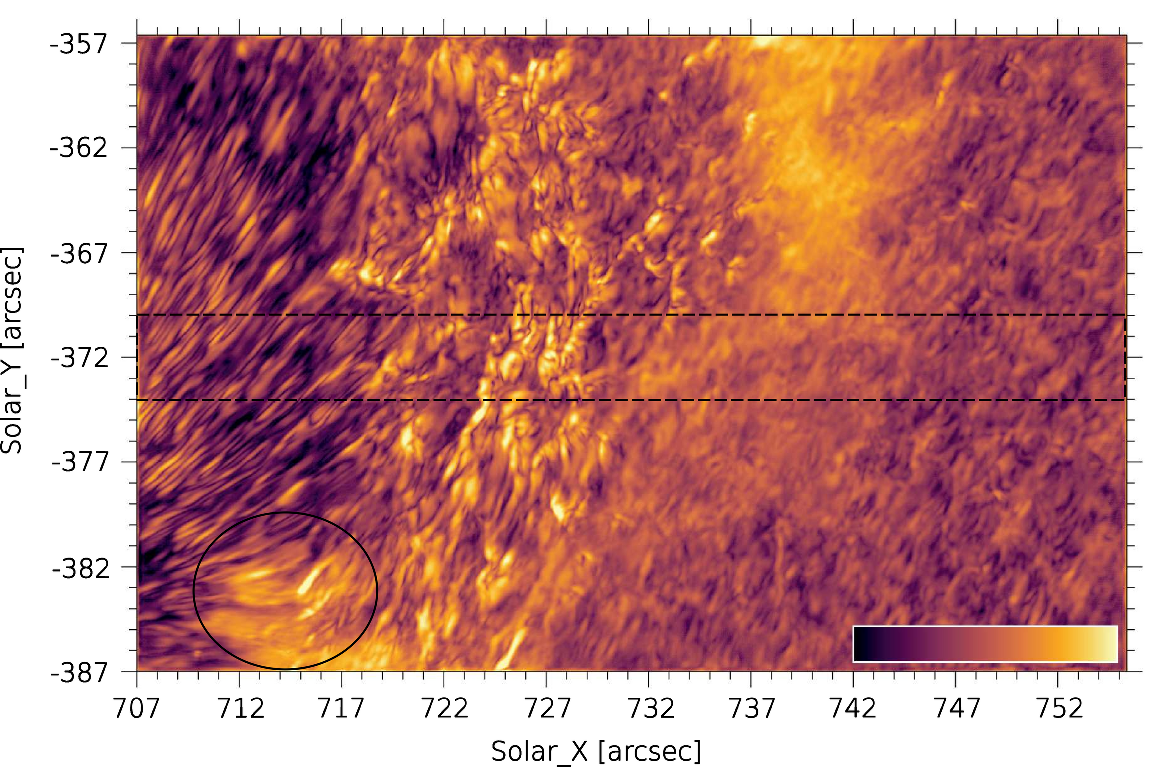}
	\caption{Logarithmic BaSAM of the \ion{Ca}{ii}~H intensity. The horizontal colour bar indicates, from left to right, increased activity. The black dashed rectangle highlights the region of the Gregor/GRIS slit-scan while the encircled area displays a region with small-scale \ion{Ca}{ii}~H brightenings (see more details in the text).}
	\label{BASAM}
\end{figure}

\begin{figure*}
	\centering
	\includegraphics[width=\textwidth]{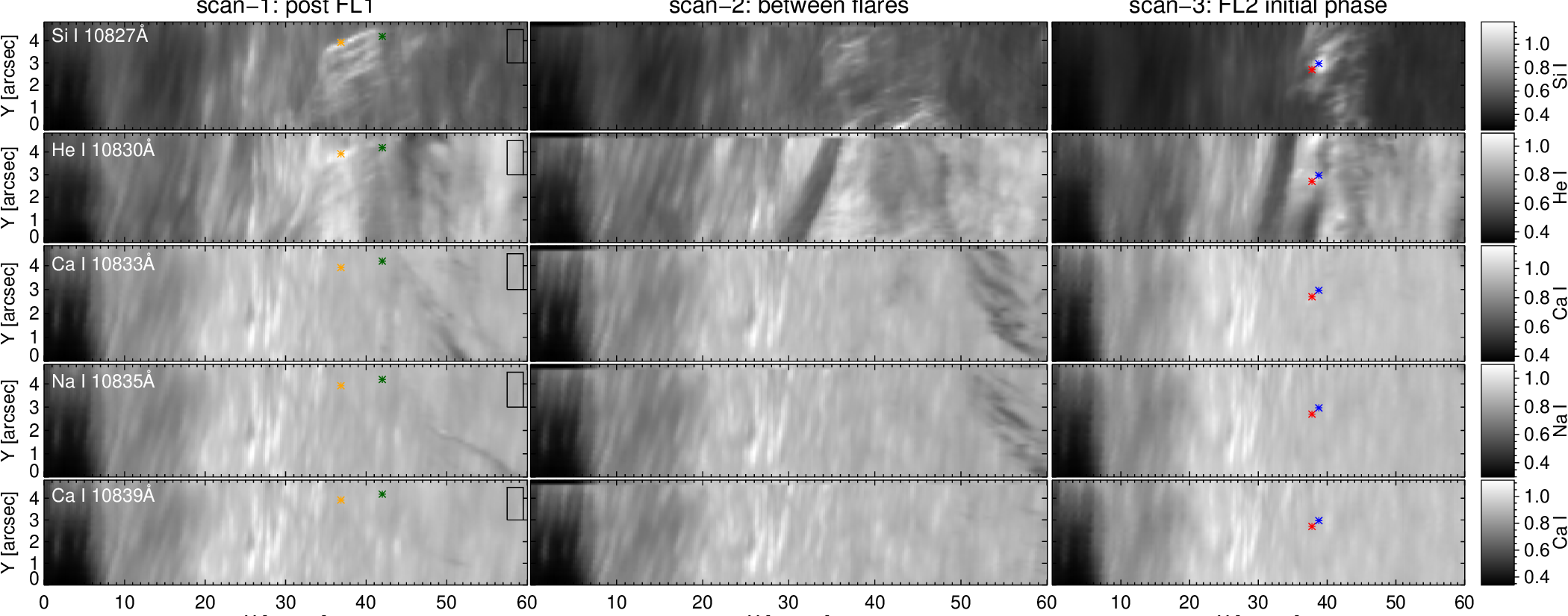}
	\caption{Spatial distribution of the intensity at the line core rest wavelength of five selected spectral lines observed with the GRIS instrument during three consecutive scans of the target region. From left to right, columns correspond to Scan 1 (08:41:53\,--\,08:44:48 UT), Scan 2 (08:45:44\,--\,08:47:43 UT), and Scan 3 (08:47:49\,--\,08:50:48 UT), respectively. From top to bottom, we show the spatial distribution of the line core rest wavelength of \ion{Si}{i}~10827~\AA, the \ion{He}{i}~10830~\AA\ triplet, \ion{Ca}{i}~10833~\AA, \ion{Na}{i}~10835~\AA, and the \ion{Ca}{i}~10839~\AA\ line. Coloured asterisks mark the location of the reference profiles that we study in detail later. The spatial region used for intensity normalization is marked by a black box in the bottom panels.}
	\label{fig:GRIS_scan}
\end{figure*}

\space

To identify the spatial location of significant changes in the solar chromosphere, Background-subtracted Solar Activity Maps (BaSAMs) were computed using HiFI+ \ion{Ca}{ii}~H images, as shown in Figure~\ref{BASAM}. BaSAMs condense the temporal variations of an entire sequence into a two-dimensional map. As explained in \citet{Denker2019}, an average two-dimensional map is calculated from the whole time series. This average is then subtracted from each individual map, and the modulus of the resulting difference maps is used to compute a second average two-dimensional map, i.e., the final BaSAM. The first application of BaSAM presented in \citet{Verma2012} and \citet{Kamlah2023} demonstrated its use on high-resolution images, while its application to spectral data is outlined in \citet{Denker2023}. The variation caused by small-scale \ion{Ca}{ii}~H brightenings can be seen near the sunspot penumbra (see the encircled area on Figure~\ref{BASAM}). The flare emission leaves a hazy appearance in the BaSAM as a patch to the west of the sunspot penumbra in a region with few bright points. This region corresponds to a quieter patch in the photospheric images, showing mostly granulation.

%$\langle | I - \langle I \rangle | \rangle$, where the angle brackets $\langle \ldots \rangle$ are shorthand for time averaging. BaSAMs capture variations in time. The flare is a transient feature moving rapidly from pixel to pixel so that the signal is washed out, whereas small-scale \ion{Ca}{ii}~H brightenings stay roughly in place so that their variations add up in place.

\begin{figure*}
\centering
\includegraphics[width=\textwidth]{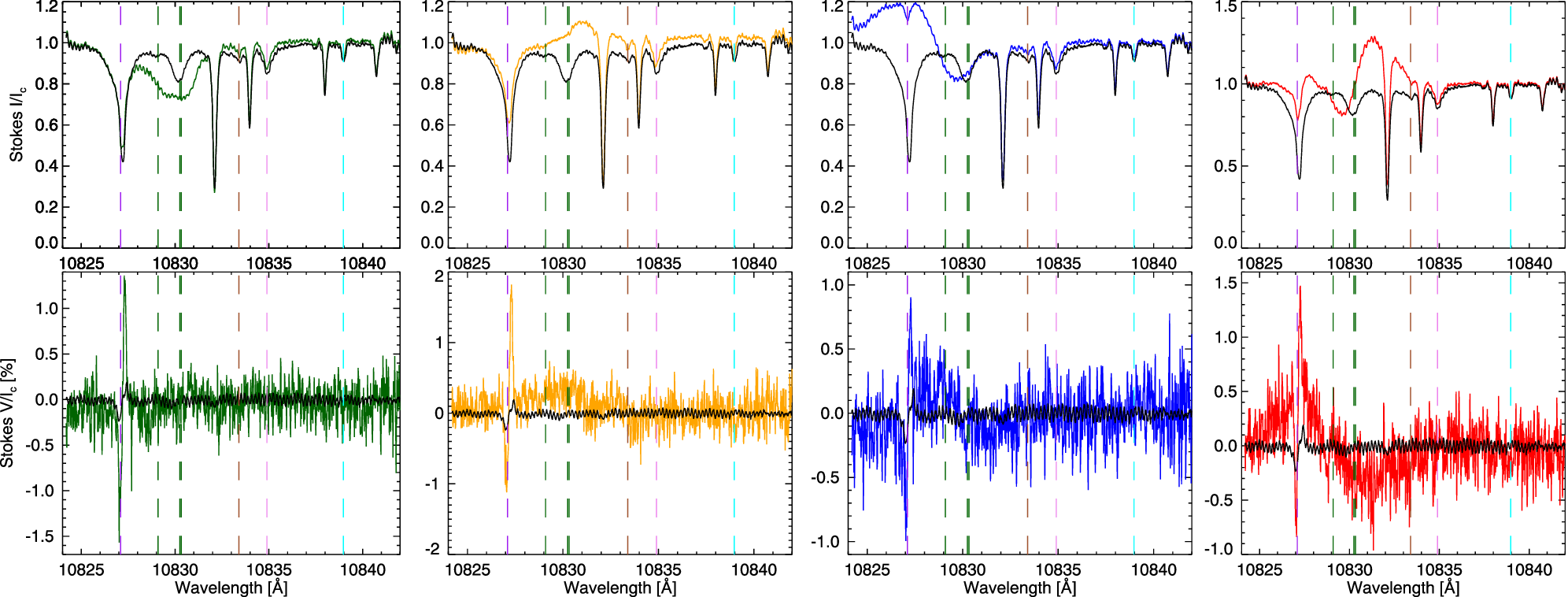}
\caption{Normalised Stokes-$I$ (top) and Stokes-$V$ (bottom) profiles from the locations marked in Figure~\ref{fig:GRIS_scan}. Vertical dashed lines indicate the rest wavelengths of \ion{Si}{i}~10827.1~\AA\ (purple), the \ion{He}{i} triplet (green), \ion{Ca}{i}~10833.4~\AA\ (brown), \ion{Na}{i}~10834.9~\AA\ (violet), and \ion{Ca}{i}~10838.9~\AA\ (cyan). The quiet-Sun region reference spectrum is shown in black.}
\label{fig:stokes_profiles}
\end{figure*}

Figure~\ref{fig:GRIS_scan} presents the temporal evolution of the GRIS observations at the line core rest wavelength of five selected spectral lines, obtained from three consecutive GRIS scans. In the case of the \ion{He}{i} spectral line, we selected as a reference wavelength that corresponds to the red component of the triplet. Each column corresponds to a different scan: Scan 1 (08:41:53\,--\,08:44:48~UT), Scan 2 (08:45:44\,--\,08:47:43~UT), and Scan 3 (08:47:49\,--\,08:50:48~UT), respectively. Rows display the spatial distribution of intensity signals for, from top to bottom, the \ion{Si}{i}~10827.1~\AA\ transition (mid-upper photosphere), the \ion{He}{i}~10830.33~\AA\ triplet (upper chromosphere), and the \ion{Ca}{i}~10833.4~\AA, \ion{Na}{i}~10834.9~\AA, and \ion{Ca}{i}~10838.9~\AA\ photospheric lines. 

The grayscale maps reveal spatial variations in the intensity between successive scans, reflecting the evolving flare-related dynamics. The most pronounced changes occur in the \ion{He}{i} triplet, consistent with its chromospheric origin, whereas more subtle intensity variations are seen in the photospheric lines. Interestingly, we can see an enhancement of the line core intensity for the silicon transition in some spatial locations. We plan to investigate more later on the possible reasons for this change, as it is somehow unexpected because the photosphere is usually mostly insensitive to flaring activity \citep[e.g.,][]{2011SSRv..159...19F}. Coloured asterisks mark specific locations within the field of view where we found complex Stokes profiles, which we aim to study in more detail.

Figure~\ref{fig:stokes_profiles} shows the Stokes profiles highlighted in Figure~\ref{fig:GRIS_scan}. We focus only on the Stokes~$I$ and $V$ profiles as linear polarisation signals are below the noise level in these particular cases. We added in black in each panel the spatially averaged profile (see squared region in the bottom panels Figure~\ref{fig:GRIS_scan}) for comparison purposes only, while colours correspond to their spatial location in Figure~\ref{fig:GRIS_scan}. 

\begin{figure*}
	\centering
	\includegraphics[width=1.00\textwidth]{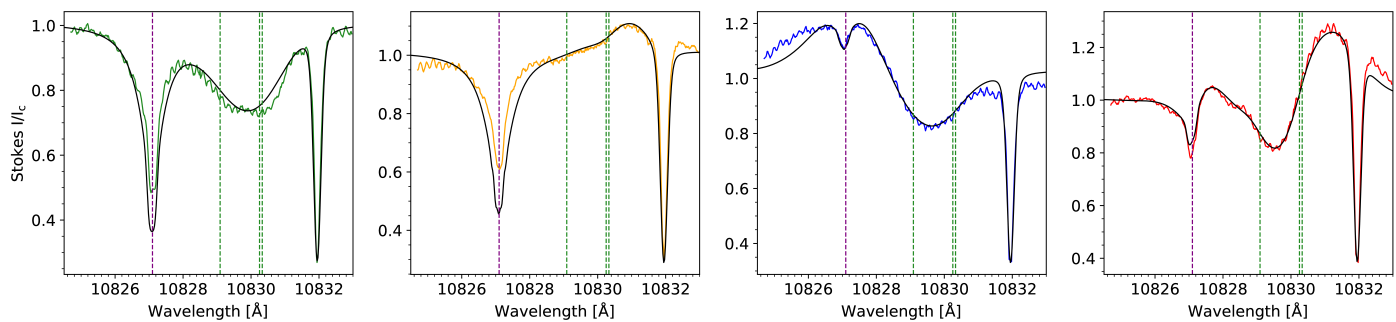}
	\caption{Comparison between the GRIS observed intensity profiles and those synthesized with the HAZEL code. Each panel corresponds to a reference profile that shows a distinct spectral shape, and whose location on the Sun is highlighted in Figure~\ref{fig:GRIS_scan}. We used the same colour code as that figure for the observations, while synthetic profiles are plotted in black. The spectral range only contains the \ion{Si}{i}~10827.1~\AA, the \ion{He}{i}~10830~\AA \ triplet, and the telluric line at 10832~\AA\ for visualization purposes. In contrast, vertical dashed lines designate the rest wavelength of the mentioned spectral lines.} 
	\label{fig:GRIS_compar}
\end{figure*}

Starting from the leftmost top panel, we can see that the \ion{He}{i} triplet intensity profile appears as a broad component (green) in comparison with the averaged profile (black) from a region with low solar activity. In the case of the next location (second from the left), the \ion{He}{i} triplet appears again as a broad profile but this time it has an enhanced intensity profile above the continuum of the spectral region (orange). In this case, we can see also a reduction in the depth of the spectral lines, particularly for the silicon transition, with respect to the reference profile. In the third selected location (third from the left), the intensity profile has become even more complex. There is one absorption contribution at the spectral range close to the rest wavelength of the \ion{He}{i} triplet, but there is a strong emission in the region close to the rest centre of the \ion{Si}{i} line (see blue). In fact, on top of this broad emission profile we can detect a weak absorption contribution that seems to be the footprint of the silicon spectral line. Finally, the rightmost intensity profile (red), corresponds to an even more complex scenario where there are multiple broad contributions, one close to the silicon transition red wing, a second close to the rest wavelength of the \ion{He}{i} triplet, and a third towards longer wavelengths, close to the telluric line at 10832~\AA.

In the case of the circular polarization signals, we can see that the \ion{Si}{i} line profiles are almost standard profiles in the first two cases where the broad contribution is far from it (see green and orange), while it becomes a bit more complex with some kind of extended wings in the other two cases (see blue and red), which could be the contribution of the multiple broad components detected in the intensity profiles.

\section{Numerical modelling} \label{sec:data}

In order to infer the information of the physical parameters from observations of the \ion{He}{i}~10830~\AA\, it is customary to use numerical codes such as the HAnle and ZEeman Light \citep[HAZEL,][]{Asensio_2008} inversion code, which takes into account the impact of atomic-level polarisation, the Zeeman, Hanle, and Paschen-Back effects when solving the radiative transfer equation and computing the emergent Stokes vector. However, as the spectra obtained during these flaring processes are so complex, we decided to follow a multi-step approach. In this study, we focus on performing synthesis with various configurations for the number of atmospheric components as well as their main physical properties, e.g., the sign or the amplitude of their LOS velocity. The goal is to produce synthetic profiles that resemble representative observed ones. This task, although manual and laborious, is fundamental to understanding what kind of phenomena is taking place. In addition, although we only focus on reproducing the intensity profiles for simplicity, this study will also help to define the baseline inversion strategy that is optimum to fit the full Stokes vector. At the same time, given the complexity of the profiles, which will require different configurations for a successful inversion, we prefer to leave the inversions for a future publication. We anticipate that a multi-step inversion process will be required where we will first classify the profiles in families with similar properties, and then invert them with tailor-made inversion configurations optimized for each family based on what we learned from the synthesis studies. Although we defer this study for a future publication, we show in the following that the simple analysis we do here will allow us to understand in detail the possible origin of the very complex profiles studied in the previous section. 

We aimed to start this trial-and-error study as simply as possible. Our initial hypothesis was that the distorted silicon spectra were due to the impact of a complex, broad, and extremely Doppler-shifted helium profile, a signature of the flare activity. Thus, we assumed that the photosphere could be represented by a reference semi-empirical 1D model which would produce standard \ion{Si}{i}~10827.1~\AA \ profiles. Therefore, the photosphere is described using the reference FALC atmosphere (we only added 300 K from $\log\tau=[-1,-3]$ to the temperature stratification) from \cite{1993ApJ...406..319F}, with a magnetic field vector of 500~G in strength, 45~degrees of inclination, and 45~degrees of azimuth constant with height for generating the \ion{Si}{i}~10827.1~\AA. After that, we added a chromosphere (a slab atmospheric model with physical parameters constant with height) that produces the \ion{He}{i}~10830~\AA\ triplet and started modifying this chromosphere until we got profiles similar to the observed ones. At this point, we did not believe the magnetic field vector played a crucial role in generating the complex intensity profiles, so we simply used a reference magnetic field with a field strength of 200~G, 45~degrees of inclination, and azimuth. Then, we modified the input values for the line-of-sight (LOS) velocity and the Doppler width to shift the location of the helium line centre and increase the width of the spectral line, respectively. Finally, we also tuned the enhancement factor $\beta$ to shift the profile shape from absorption to emission (see more details in the HAZEL manual\footnote{\href{https://aasensio.github.io/hazel2/index.html}{aasensio.github.io/hazel2/}}), leaving the rest of atmospheric parameters with the default values.

We present in Figure~\ref{fig:GRIS_compar} the observed (colours) and synthetic (black) intensity profiles of a spectral region containing only the \ion{Si}{i}~10827.1~\AA\ line, the \ion{He}{i}~10830~\AA\ triplet, and the telluric line at 10832~\AA. The first two profiles from the left are fitted with only one chromospheric component, while the third and fourth profiles from the left required two and three chromospheric components, respectively. Interestingly, all profiles share some properties. First, and most importantly, all \ion{He}{i} triplet profiles exhibit a broad profile covering a wide range of wavelengths, making it impossible to distinguish the blue and the red components of the triplet. We reproduced this effect by increasing the Doppler width of the lines, with values in the range 15\,--\,25 km s$^{-1}$. Such high values indicate that the spectral line is being formed in atmospheric layers where the plasma has either high temperature or strong turbulent velocities are present in the resolution element. Additionally, the single chromospheric component used in the first two profiles from the left is very different, requiring absorption in the leftmost profile and emission in the other case. The LOS velocities for the chromospheric components used to match each profile are $-10$ and 18 km s$^{-1}$, respectively (where negative means blueshifts). Those values are not extreme for the case of the \ion{He}{i} triplet in a flaring scenario \citep[e.g. ][]{2011SSRv..159...19F, Kuckein2025}. 

In the case of the third and fourth profiles from the left, their shape is more complex, requiring additional chromospheric components. For instance, in the case of the third panel, we have a strong emission near the \ion{Si}{i}~10827.1~\AA \ rest wavelength and then an absorption near the 10830~\AA \ region. We decided to add two chromospheres, similar to the individual ones from the two previous locations, i.e., one in absorption (used in the first profile) and the second one in emission (used on the second profile from the left). The component producing the absorption is also wide, so we used a Doppler width of 25 km s$^{-1}$. The centre of the triplet is blueshifted with a LOS velocity of $-19$ km s$^{-1}$. The chromospheric component producing the emission spectral feature uses the same Doppler width but is blueshifted with a LOS velocity of $-88$ km s$^{-1}$. Interestingly, a good fit of the \ion{Si}{i}~10827.1~\AA \ line is obtained without any significant modification of the FALC atmosphere except for the mentioned slight increase of 300 K between $\log \tau_{500}=[-2,-3]$. Thus, we can conclude that the main element producing the complex spectral features found between [10824,10828]~\AA \ is a strong and broad helium component that is superimposed on an almost standard silicon profile. Finally, the profile in the rightmost panel is just a continuation of building complexity from the previous cases. We can see three distinct broad peaks, two in emission and one in absorption, so we proceeded to use three chromospheric components similar to the previous ones, centred at $-78$, $-28$, and $+25$ km s$^{-1}$ (where the negative sign indicates blueshifted motions).

\section{Discussion}

We have developed a strategy to accurately reproduce a selection of the most complex profiles observed. In the first two panels from the left in Figure~\ref{fig:GRIS_compar}, we can infer that there is a main contribution of hot, rather turbulent, chromospheric plasma moving at a certain speed along the LOS. The main difference between the two panels from the left is possibly a different temperature, higher for the second profile from the left, which results in a \ion{He}{i} profile with a higher intensity than that of the local continuum. In the case of the two rightmost panels, we have an additional ingredient of complexity because, besides having broad profiles, we also have multiple broad components, some of them close to rest velocities and others showing extreme Doppler shift values. The reason for that seems to be the juxtaposition of numerous flare phases. We believe we have the contribution of the decaying phase from the first flare, specifically the broad component close to the rest wavelength of helium, along with a high-speed component created by the impulsive phase of the second flaring process. A more detailed physical interpretation of the events is beyond the scope of this work. It would require the detailed inversion of the Stokes profiles and analysing the spatial and temporal characteristics of the inferred properties. Despite that, we can use the four cases studied here to understand some general properties of the flaring process. In general, we find large Doppler widths and some chromospheric components in emission. If the large Doppler widths (15\,--\,25 km s$^{-1}$) are caused by thermal broadening, we can infer the temperature using
\begin{equation}
\Delta v = \sqrt{\frac{2 k_B T }{m}},
\end{equation}
where $k_B$ is the Boltzmann constant, $T$ the thermal temperature, and $m$ the mass of helium. The results indicate that we would need a temperature ranging from around 50000 to 150000 K. Those temperatures are too high for the region of formation of the \ion{He}{i} spectral line triplet, so we conclude that non-thermal quasi-random motions are playing a significant role in the flaring region. The presence of such turbulent motions is concomitant with the presence of strong unidirectional flows, as indicated by the LOS velocities inferred from the synthesis.

The presence of emission in the \ion{He}{i}~10830~\AA\ triplet is not standard in on-disc solar observations. It is modelled in HAZEL as an enhancement factor $\beta$ that modifies the source function of the slab where the \ion{He}{i} triplet is:
\begin{equation}
\mathbf{I} = e^{-{\bf K}\tau} {\mathbf I}_{\rm sun} + \mathbf{K}^{-1} \left( 1-e^{-\mathbf{K}\tau} \right) \beta \mathbf S,
\end{equation} 
where $\mathbf{I}_{\rm sun}$ is the Stokes vector that illuminates the slab \citep[see][for more information]{Asensio_2008}, $\tau$ is the optical depth on the red component of the triplet, $ \textbf{K}$ the propagation matrix, and $ \textbf{S}$ is the source function. Although it is not straightforward to provide an actual physical meaning to the parameter $\beta$, we can understand it as an overpopulation factor of the $^3P$ upper level of the 10830~\AA \ triplet. This overpopulation is surely produced by overionisation produced by the flaring process, which, after the ladder of recombinations, leads to a trapping of population in the $^3P$ level.

Also, as part of future work, we need to disentangle the meaning of the inferred LOS velocities. Since we are very close to the limb, blue- or redshifted motions with respect to the observer need to be projected back into the local reference frame so we may be able to get an accurate estimation of the real plasma velocities and orientation during the flaring activity.

Finally, we cannot forget that we observed the four Stokes parameters, and hence we should also try to infer and interpret the magnetic properties of these multiple chromospheric components. It is even more challenging than fitting Stokes $I$ because their signals are weak. This is a consequence, partially, of the fact that Stokes $I$ profiles are very broad. Additionally, the presence of several components complicates the shape of the polarimetric profiles, making it more challenging to interpret the signals.

\section{Conclusion} \label{sec:conclusion}

We presented full-Stokes spectropolarimetry and high-resolution imaging of observations of the dynamic evolution of the photosphere and chromosphere during and after two C-class flares (FL1 and FL2) in active region NOAA~13363. We analysed slit-scan spectra in the infrared spectral region around the \ion{He}{i} triplet, including the photospheric \ion{Na}{i}, \ion{Ca}{i}, and upper-photospheric \ion{Si}{i} lines, using the GRIS infrared spectrograph. The photospheric lines did not show significant changes during the flare events, consistent with their photospheric formation. In contrast, the upper chromospheric \ion{He}{i} triplet exhibited a clear increase in intensity starting with FL1, sometimes transitioning into emission. To quantify these changes, we computed background-subtracted activity maps (BaSAMs) from HiFI+ speckle-restored \ion{Ca}{ii}~H images, which revealed two distinct flare components: a diffuse haze related to FL2 and narrow and bright filaments close to the active region. Time-lapse analysis indicates that the flare emission did not significantly affect the inverse granulation or nearby plage. The filament morphology suggests a link to magnetic field reconfiguration, while disturbed fine-scale emission hints at plasma turbulence.

The enhanced \ion{He}{i} emission we observe is consistent with previous studies. For example, intensity ratios of \ion{He}{i} to continuum of 1.36, 1.30, and 1.15 have been reported during M2.0, C9.7, and C2.0 flares, respectively \citep{Li2007,Penn1995,Sasso2011}, while \citet{Kuckein2015} reported ratios of about 1.86 in an M3.2 flare. In our observations, the \ion{He}{i} profiles exhibit strong red- and blueshifted Doppler shifts close to $-90$~km~s$^{-1}$. These velocities are supersonic, given the local sound speed of $\sim$10~km~s$^{-1}$ at the formation temperature of 8000\,--\,10000~K, and are consistent with field-aligned evaporation and condensation flows commonly reported in flaring regions or associated with filament eruptions \citep{Li2007, Kuckein2015, Sasso2011, refId0, Sowmya2022}.

We also complemented the analysis of the spectral features by doing forward modelling with the HAZEL code. Our theoretical studies reveal that the different broad components, either in emission or absorption, redshifted or blueshifted, can be reproduced with a helium contribution with a large Doppler width. The larger the number of contributions, the larger the number of chromospheric components we needed to add to the synthesis configuration. Interestingly, we confirmed that the apparent enhancement in the \ion{Si}{i} spectral range seen in some cases is simply due to the overlap with a strongly blueshifted \ion{He}{i} emission component, rather than intrinsic photospheric heating \citep{Judge_2015}.

Our observations confirm the chromospheric origin of the \ion{He}{i} emission and show that the extreme red- and blueshifted profiles reflect a complex response of the flaring chromosphere. Radiative excitation from coronal EUV irradiation, energy deposition by flare-accelerated electrons, and dynamic field-aligned plasma flows likely act together to drive the observed supersonic downflows and upflows. In contrast, the photospheric lines (\ion{Ca}{i}, \ion{Na}{i}, and mostly \ion{Si}{i}) show no measurable changes in their cores, demonstrating that the photosphere (at least the lower part) is not directly heated by flare energy deposition.

Multi-line diagnostics such as those presented here are therefore crucial to separate chromospheric and photospheric flare responses. However, we plan to delve more into the analysis of these data through inversions of the multiple observed maps. We aim to infer the spatial distribution and evolution of atmospheric parameters across the entire dataset. By doing so, we strive to constrain the role of different physical mechanisms, such as conduction, ionisation, and indirect coupling, in transporting energy from upper layers into the lower solar atmosphere.

\begin{acknowledgements}
 The 1.5-meter Gregor solar telescope was built by a German consortium under the leadership of the Institute for Solar Physics (KIS) in Freiburg with the Leibniz Institute for Astrophysics Potsdam (AIP), the Institute for Astrophysics G\"ottingen (IAG), and the Max Planck Institute for Solar System Research (MPS) in G\"ottingen as partners, and with contributions by the Instituto de Astrof\'isica de Canarias and the Astronomical Institute of the Academy of Sciences of the Czech Republic. The Gregor AO and instrument distribution optics were redesigned by KIS, whose technical staff is gratefully acknowledged. This research has received financial support from the European Union's Horizon 2020 research and innovation program under grant agreement No. 824135 (SOLARNET). Work acknowledges grant RYC2022-037660 funded by MCIN/AEI/10.13039/501100011033 and by ESF Investing in your future. This work was supported by the Shota Rustaveli National Science Foundation (SRNSF) grant YS-22-407. This research was made possible by the Fellowship for Excellent Researchers R2-R4 (09I03-03-V04-00015). CQN acknowledges support from the Agencia Estatal de Investigación del Ministerio de Ciencia, Innovación y Universidades (MCIU/AEI) under grant ``Polarimetric Inference of Magnetic Fields'' and the European Regional Development Fund (ERDF) with reference PID2022-136563NB-I00/10.13039/501100011033. The publication is part of Project ICTS2022-007828, funded by MICIN and the European Union NextGenerationEU/RTRP.  CQN also acknowledges support from Grants PID2024-156538NB-I00 and PID2024-156066OB-C55 funded by MCIN/AEI/10.13039/501100011033 and by ``ERDF A way of making Europe''. MB, PG and JR acknowledge the support of the project VEGA 2/0043/24. MV acknowledges the support from IGSTC-WISER grant (IGSTC-05373).
 SL acknowledges the support of Štefan Schwarz Support Fund 2023/OV1/015.
\end{acknowledgements}

\bibliographystyle{aa}  % A&A's required bibliography style
\bibliography{References.bib} 

\begin{thebibliography}{60}
\expandafter\ifx\csname natexlab\endcsname\relax\def\natexlab#1{#1}\fi

\bibitem[{{Akimov} {et~al.}(2014){Akimov}, {Belkina}, \&
  {Marchenko}}]{Akimov2014}
{Akimov}, L.~A., {Belkina}, I.~L., \& {Marchenko}, G.~P. 2014, MNRAS, 439, 193

\bibitem[{Allred {et~al.}(2015)Allred, Kowalski, \&
  Carlsson}]{Allred2015ApJ809}
Allred, J.~C., Kowalski, A.~F., \& Carlsson, M. 2015, ApJ, 809, 104

\bibitem[{Anan {et~al.}(2018)Anan, Yoneya, Ichimoto, UeNo, Shiota, Nozawa,
  Takasao, \& Kawate}]{tetsu2018}
Anan, T., Yoneya, T., Ichimoto, K., {et~al.} 2018, PASJ, 70, 101

\bibitem[{Aschwanden(2005)}]{Aschwanden2005}
Aschwanden, M.~J. 2005, Physics of the Solar Corona, Springer Praxis Books
  (Dordrecht: Springer Berlin, Heidelberg), XXII, 908

\bibitem[{{Asensio Ramos, A.} {et~al.}(2008){Asensio Ramos, A.}, {Trujillo
  Bueno J.}, \& {Landi Degl’Innocenti E.}}]{Asensio_2008}
{Asensio Ramos, A.}, {Trujillo Bueno J.}, \& {Landi Degl’Innocenti E.} 2008,
  ApJ, 683, 542

\bibitem[{{Balthasar} {et~al.}(2016){Balthasar}, {G{\"o}m{\"o}ry},
  {Gonz{\'a}lez Manrique}, {Kuckein}, {Kavka}, {Ku{\v c}era}, {Schwartz},
  {Va{\v s}kov{\'a}}, {Berkefeld}, {Collados Vera}, {Denker}, {Feller},
  {Hofmann}, {Lagg}, {Nicklas}, {Orozco Su{\'a}rez}, {Pastor Yabar}, {Rezaei},
  {Schlichenmaier}, {Schmidt}, {Schmidt}, {Sigwarth}, {Sobotka}, {Solanki},
  {Soltau}, {Staude}, {Strassmeier}, {Volkmer}, {von der L{\"u}he}, \&
  {Waldmann}}]{balthasar2016}
{Balthasar}, H., {G{\"o}m{\"o}ry}, P., {Gonz{\'a}lez Manrique}, S.~J., {et~al.}
  2016, AN, 337, 1050

\bibitem[{{Berkefeld } {et~al.}(2012){Berkefeld }, {Schmidt}, {Soltau}, {von
  der L{\"u}he}, \& {Heidecke}}]{berkefeld2012}
{Berkefeld }, T., {Schmidt}, D., {Soltau}, D., {von der L{\"u}he}, O., \&
  {Heidecke}, F. 2012, {Astron.\ Nachr.}, 333, 863

\bibitem[{{Berkefeld} {et~al.}(2010){Berkefeld}, {Soltau}, {Schmidt}, \& {von
  der L\"{u}he}}]{berkefeld2010}
{Berkefeld}, T., {Soltau}, D., {Schmidt}, D., \& {von der L\"{u}he}, O. 2010,
  Appl. Opt., 49, G155

\bibitem[{Chen {et~al.}(2024)Chen, Bastian, \& Gary}]{Chen2024}
Chen, B., Bastian, T.~S., \& Gary, D.~E. 2024, ApJ, 971, 128

\bibitem[{{Collados}(1999)}]{collados1999}
{Collados}, M. 1999, in ASPC, Vol. 184, Third Advances in Solar Physics
  Euroconference: Magnetic Fields and Oscillations, ed. B.~{Schmieder},
  A.~{Hofmann}, \& J.~{Staude}, 3--22

\bibitem[{{Collados} {et~al.}(2012){Collados}, {L{\'o}pez}, {P{\'a}ez},
  {Hern{\'a}ndez}, {Reyes}, {Calcines}, {Ballesteros}, {D{\'{\i}}az}, {Denker},
  {Lagg}, {Schlichenmaier}, {Schmidt}, {Solanki}, {Strassmeier}, {von der
  L{\"u}he}, \& {Volkmer}}]{collados2012}
{Collados}, M., {L{\'o}pez}, R., {P{\'a}ez}, E., {et~al.} 2012, AN, 333, 872

\bibitem[{Collados(2003)}]{collados2003}
Collados, M.~V. 2003, in Proc.\ SPIE, Vol. 4843, Polarimetry in Astronomy, ed.
  S.~Fineschi, 55 -- 65

\bibitem[{Denker {et~al.}(2018{\natexlab{a}})Denker, Dineva, Balthasar, Verma,
  Kuckein, Diercke, \& Manrique}]{denker_2018}
Denker, C., Dineva, E., Balthasar, H., {et~al.} 2018{\natexlab{a}}, SoPh, 293,
  44

\bibitem[{Denker {et~al.}(2018{\natexlab{b}})Denker, Kuckein, Verma, Manrique,
  Diercke, Enke, Klar, Balthasar, Louis, \& Dineva}]{Denker2018}
Denker, C., Kuckein, C., Verma, M., {et~al.} 2018{\natexlab{b}}, ApJSS, 236, 5

\bibitem[{{Denker} \& {Verma}(2019)}]{Denker2019}
{Denker}, C. \& {Verma}, M. 2019, SoPh, 294, 71

\bibitem[{Denker {et~al.}(2023)Denker, Verma, Wiśniewska, Kamlah,
  Kontogiannis, Dineva, Rendtel, Bauer, Dionies, {\"O}nel, Woche, Kuckein,
  Seelemann, \& Pal}]{Denker2023}
Denker, C., Verma, M., Wiśniewska, A., {et~al.} 2023, JATIS, 9, 015001

\bibitem[{Ding {et~al.}(2005)Ding, Li, \& Fang}]{Ding2005AandA432}
Ding, M.~D., Li, H., \& Fang, C. 2005, A\&A, 432, 699

\bibitem[{{Fisher} {et~al.}(1985){Fisher}, {Canfield}, \&
  {McClymont}}]{Fisher1985}
{Fisher}, G.~H., {Canfield}, R.~C., \& {McClymont}, A.~N. 1985, \apj, 289, 414

\bibitem[{Fisher(1964)}]{Fisher1964}
Fisher, R.~R. 1964, Astrophysical Journal, 140, 1326

\bibitem[{{Fletcher} {et~al.}(2011){Fletcher}, {Dennis}, {Hudson}, {Krucker},
  {Phillips}, {Veronig}, {Battaglia}, {Bone}, {Caspi}, {Chen}, {Gallagher},
  {Grigis}, {Ji}, {Liu}, {Milligan}, \& {Temmer}}]{2011SSRv..159...19F}
{Fletcher}, L., {Dennis}, B.~R., {Hudson}, H.~S., {et~al.} 2011, \ssr, 159, 19

\bibitem[{{Fontenla} {et~al.}(1993){Fontenla}, {Avrett}, \&
  {Loeser}}]{1993ApJ...406..319F}
{Fontenla}, J.~M., {Avrett}, E.~H., \& {Loeser}, R. 1993, \apj, 406, 319

\bibitem[{Graham \& Cauzzi(2015)}]{Graham_2015}
Graham, D.~R. \& Cauzzi, G. 2015, ApJL, 807, L22

\bibitem[{{Hirayama}(1974)}]{Hirayama1974}
{Hirayama}, T. 1974, \solphys, 34, 323

\bibitem[{Hudson \& Ryan(1995)}]{HudsonRyan1995}
Hudson, H.~S. \& Ryan, J.~M. 1995, Annual Review of Astronomy and Astrophysics,
  33, 239

\bibitem[{{Joshi} {et~al.}(2025){Joshi}, {Dud{\'\i}k}, {Schmieder}, {Aulanier},
  \& {Chandra}}]{Joshi2025}
{Joshi}, R., {Dud{\'\i}k}, J., {Schmieder}, B., {Aulanier}, G., \& {Chandra},
  R. 2025, \aap, 698, A301

\bibitem[{Judge {et~al.}(2015)Judge, Kleint, \& Dalda}]{Judge_2015}
Judge, P.~G., Kleint, L., \& Dalda, A.~S. 2015, ApJ, 814, 100

\bibitem[{{Kamlah} {et~al.}(2023){Kamlah}, {Verma}, {Denker}, \&
  {Wang}}]{Kamlah2023}
{Kamlah}, R., {Verma}, M., {Denker}, C., \& {Wang}, H. 2023, A\&A, 675, A182

\bibitem[{{Kleint} {et~al.}(2020){Kleint}, {Berkefeld}, {Esteves}, {Sonner},
  {Volkmer}, {Gerber}, {Kr{\"a}mer}, {Grassin}, \& {Berdyugina}}]{Kleint2020}
{Kleint}, L., {Berkefeld}, T., {Esteves}, M., {et~al.} 2020, \aap, 641, A27

\bibitem[{Kuckein {et~al.}(2015)Kuckein, Collados, \& Sainz}]{Kuckein2015}
Kuckein, C., Collados, M., \& Sainz, R.~M. 2015, ApJL, 799, L25

\bibitem[{{Kuckein} {et~al.}(2017){Kuckein}, {Denker}, {Verma}, {Balthasar},
  {Gonz\'{a}lez Manrique}, {Louis}, \& {Diercke}}]{stools}
{Kuckein}, C., {Denker}, C., {Verma}, M., {et~al.} 2017, in IAUS, Vol. 327,
  Fine Structure and Dynamics of the Solar Atmosphere, ed. S.~{Vargas
  Dom{\'\i}nguez}, A.~G. {Kosovichev}, P.~{Antolin}, \& L.~{Harra}, 20--24

\bibitem[{{Kuckein, C.} {et~al.}(2025){Kuckein, C.}, {Collados, M.}, {Asensio
  Ramos, A.}, {Díaz Baso, C. J.}, {Felipe, T.}, {Quintero Noda, C.}, {Kleint,
  L.}, {Fletcher, L.}, \& {Matthews, S.}}]{Kuckein2025}
{Kuckein, C.}, {Collados, M.}, {Asensio Ramos, A.}, {et~al.} 2025, A\&A, 699,
  A121

\bibitem[{{Kuckein, C.} {et~al.}(2020){Kuckein, C.}, {González Manrique, S.
  J.}, {Kleint, L.}, \& {Asensio Ramos, A.}}]{refId0}
{Kuckein, C.}, {González Manrique, S. J.}, {Kleint, L.}, \& {Asensio Ramos,
  A.} 2020, A\&A, 640, A71

\bibitem[{{Lemen} {et~al.}(2012){Lemen}, {Title}, {Akin}, {Boerner}, {Chou},
  {Drake}, {Duncan}, {Edwards}, {Friedlaender}, {Heyman}, {Hurlburt}, {Katz},
  {Kushner}, {Levay}, W., P., L., Sarah., A., J., A., A., D., Jean-Pierre.,
  Jacob., Carl., A., N., David., E., Richard., Leon., Sang., A., I., H., A.,
  Peter., Gary., Paul., Peter., Regina., L., Sarah., Matthew., James., \&
  Nicholas.}]{Lemen2012}
{Lemen}, J.~R., {Title}, A.~M., {Akin}, D.~J., {et~al.} 2012, SoPh, 275, 17

\bibitem[{Li {et~al.}(2022)Li, Fang, Li, Ding, Chen, Qiu, You, Yuan, An, Tao,
  Li, Chen, Liu, Mei, Liang, Zhang, Cheng, Chen, Chen, \& Zhao}]{Li_2022}
Li, C., Fang, C., Li, Z., {et~al.} 2022, Science China: Physics, Mechanics and
  Astronomy

\bibitem[{Li {et~al.}(2019)Li, Chen, Feng, Li, Huang, Li, Lu, Xue, Ying, Zhao,
  Yang, Gan, Fang, Song, Wang, Guo, He, Zhu, Zhu, Deng, Bao, Cao, \&
  Yang}]{Li_2019}
Li, H., Chen, B., Feng, L., {et~al.} 2019, Research in Astronomy and
  Astrophysics, 19, 158

\bibitem[{Li {et~al.}(2007)Li, You, Yu, \& Du}]{Li2007}
Li, H., You, J., Yu, X., \& Du, Q. 2007, SoPh, 241, 301

\bibitem[{Li {et~al.}(2024)Li, Zhang, Chen, \& Wang}]{Li2024}
Li, X., Zhang, J., Chen, Y., \& Wang, H. 2024, SoPh, 299, 85

\bibitem[{{Nagai} \& {Emslie}(1984)}]{nagai1984}
{Nagai}, F. \& {Emslie}, A.~G. 1984, ApJ, 279, 896

\bibitem[{{Neupert}(1968)}]{Neupert1968}
{Neupert}, W.~M. 1968, \apjl, 153, L59

\bibitem[{{Penn} \& {Kuhn}(1995)}]{Penn1995}
{Penn}, M.~J. \& {Kuhn}, J.~R. 1995, ApJL, 441, 51

\bibitem[{{Pesnell} {et~al.}(2012){Pesnell}, {Thompson}, \&
  {Chamberlin}}]{Pesnell2012}
{Pesnell}, W.~D., {Thompson}, B.~J., \& {Chamberlin}, P.~C. 2012, SoPh, 275, 3

\bibitem[{Petrie \& Sudol(2010)}]{petrie2010}
Petrie, G. J.~D. \& Sudol, J.~J. 2010, ApJ, 724, 1218

\bibitem[{{Quintero Noda} {et~al.}(2022){Quintero Noda}, {Schlichenmaier},
  {Bellot Rubio}, {L{\"o}fdahl}, {Khomenko}, {Jur{\v{c}}{\'a}k}, {Leenaarts},
  {Kuckein}, {Gonz{\'a}lez Manrique}, {Gun{\'a}r}, {Nelson}, {de la Cruz
  Rodr{\'\i}guez}, {Tziotziou}, {Tsiropoula}, {Aulanier}, {Aboudarham},
  {Allegri}, {Alsina Ballester}, {Amans}, {Asensio Ramos}, {Bail{\'e}n},
  {Balaguer}, {Baldini}, {Balthasar}, {Barata}, {Barczynski}, {Barreto
  Cabrera}, {Baur}, {B{\'e}chet}, {Beck}, {Bel{\'\i}o-As{\'\i}n},
  {Bello-Gonz{\'a}lez}, {Belluzzi}, {Bentley}, {Berdyugina}, {Berghmans},
  {Berlicki}, {Berrilli}, {Berkefeld}, {Bettonvil}, {Bianda}, {Bienes
  P{\'e}rez}, {Bonaque-Gonz{\'a}lez}, {Braj{\v{s}}a}, {Bommier}, {Bourdin},
  {Burgos Mart{\'\i}n}, {Calchetti}, {Calcines}, {Calvo Tovar}, {Campbell},
  {Carballo-Mart{\'\i}n}, {Carbone}, {Carlin}, {Carlsson}, {Castro L{\'o}pez},
  {Cavaller}, {Cavallini}, {Cauzzi}, {Cecconi}, {Chulani}, {Cirami},
  {Consolini}, {Coretti}, {Cosentino}, {C{\'o}zar-Castellano}, {Dalmasse},
  {Danilovic}, {De Juan Ovelar}, {Del Moro}, {del Pino Alem{\'a}n}, {del Toro
  Iniesta}, {Denker}, {Dhara}, {Di Marcantonio}, {D{\'\i}az Baso}, {Diercke},
  {Dineva}, {D{\'\i}az-Garc{\'\i}a}, {Doerr}, {Doyle}, {Erdelyi}, {Ermolli},
  {Escobar Rodr{\'\i}guez}, {Esteban Pozuelo}, {Faurobert}, {Felipe}, {Feller},
  {Feijoo Amoedo}, {Femen{\'\i}a Castell{\'a}}, {Fernandes}, {Ferro
  Rodr{\'\i}guez}, {Figueroa}, {Fletcher}, {Franco Ordovas}, {Gafeira},
  {Gardenghi}, {Gelly}, {Giorgi}, {Gisler}, {Giovannelli}, {Gonz{\'a}lez},
  {Gonz{\'a}lez}, {Gonz{\'a}lez-Cava}, {Gonz{\'a}lez Garc{\'\i}a},
  {G{\"o}m{\"o}ry}, {Gracia}, {Grauf}, {Greco}, {Grivel}, {Guerreiro},
  {Guglielmino}, {Hammerschlag}, {Hanslmeier}, {Hansteen}, {Heinzel},
  {Hern{\'a}ndez-Delgado}, {Hern{\'a}ndez Su{\'a}rez}, {Hidalgo}, {Hill},
  {Hizberger}, {Hofmeister}, {J{\"a}gers}, {Janett}, {Jarolim}, {Jess},
  {Jim{\'e}nez Mej{\'\i}as}, {Jolissaint}, {Kamlah}, {Kapit{\'a}n},
  {Ka{\v{s}}parov{\'a}}, {Keller}, {Kentischer}, {Kiselman}, {Kleint},
  {Klvana}, {Kontogiannis}, {Krishnappa}, {Ku{\v{c}}era}, {Labrosse}, {Lagg},
  {Landi Degl'Innocenti}, {Langlois}, {Lafon}, {Laforgue}, {Le Men}, {Lepori},
  {Lepreti}, {Lindberg}, {Lilje}, {L{\'o}pez Ariste}, {L{\'o}pez
  Fern{\'a}ndez}, {L{\'o}pez Jim{\'e}nez}, {L{\'o}pez L{\'o}pez}, {Manso
  Sainz}, {Marassi}, {Marco de la Rosa}, {Marino}, {Marrero}, {Mart{\'\i}n},
  {Mart{\'\i}n G{\'a}lvez}, {Mart{\'\i}n Hernando}, {Masciadri}, {Mart{\'\i}nez
  Gonz{\'a}lez}, {Matta-G{\'o}mez}, {Mato}, {Mathioudakis}, {Matthews}, {Mein},
  {Merlos Garc{\'\i}a}, {Moity}, {Montilla}, {Molinaro}, {Molodij}, {Montoya},
  {Munari}, {Murabito}, {N{\'u}{\~n}ez Cagigal}, {Oliviero}, {Orozco
  Su{\'a}rez}, {Ortiz}, {Padilla-Hern{\'a}ndez}, {Pa{\'e}z Ma{\~n}{\'a}},
  {Paletou}, {Pancorbo}, {Pastor Ca{\~n}edo}, {Pastor Yabar}, {Peat},
  {Pedichini}, {Peixinho}, {Pe{\~n}ate}, {P{\'e}rez de Taoro}, {Peter},
  {Petrovay}, {Piazzesi}, {Pietropaolo}, {Pleier}, {Poedts}, {P{\"o}tzi},
  {Podladchikova}, {Prieto}, {Quintero Nehrkorn}, {Ramelli}, {Ramos Sapena},
  {Rasilla}, {Reardon}, {Rebolo}, {Regalado Olivares}, {Reyes
  Garc{\'\i}a-Talavera}, {Riethm{\"u}ller}, {Rimmele}, {Rodr{\'\i}guez
  Delgado}, {Rodr{\'\i}guez Gonz{\'a}lez}, {Rodr{\'\i}guez-Losada},
  {Rodr{\'\i}guez Ramos}, {Romano}, {Roth}, {Rouppe van der Voort}, {Rudawy},
  {Ruiz de Galarreta}, {Ryb{\'a}k}, {Salvade}, {S{\'a}nchez-Capuchino},
  {S{\'a}nchez Rodr{\'\i}guez}, {Sangiorgi}, {Say{\`e}de}, {Scharmer},
  {Scheiffelen}, {Schmidt}, {Schmieder}, {Scir{\`e}}, {Scuderi}, {Siegel},
  {Sigwarth}, {Sim{\~o}es}, {Snik}, {Sliepen}, {Sobotka}, {Socas-Navarro},
  {Sola La Serna}, {Solanki}, {Soler Trujillo}, {Soltau}, {Sordini}, {Sosa
  M{\'e}ndez}, {Stangalini}, {Steiner}, {Stenflo}, {{\v{S}}t{\v{e}}p{\'a}n},
  {Strassmeier}, {Sudar}, {Suematsu}, {S{\"u}tterlin}, {Tallon}, {Temmer},
  {Tenegi}, {Tritschler}, {Trujillo Bueno}, {Turchi}, {Utz}, {van Harten}, {van
  Noort}, {van Werkhoven}, {Vansintjan}, {Vaz Cedillo}, {Vega Reyes}, {Verma},
  {Veronig}, {Viavattene}, {Vitas}, {V{\"o}gler}, {von der L{\"u}he},
  {Volkmer}, {Waldmann}, {Walton}, {Wisniewska}, {Zeman}, {Zeuner}, {Zhang},
  {Zuccarello}, \& {Collados}}]{QuinteroNoda2022EST}
{Quintero Noda}, C., {Schlichenmaier}, R., {Bellot Rubio}, L.~R., {et~al.}
  2022, \aap, 666, A21

\bibitem[{Reames(2013)}]{Reames2013}
Reames, D.~V. 2013, Space Sci. Rev., 175, 53

\bibitem[{Reep {et~al.}(2016)Reep, Warren, Crump, \& Simões}]{Reep2016ApJ827}
Reep, J.~W., Warren, H.~P., Crump, N.~A., \& Simões, P.~J.~A. 2016, ApJ, 827,
  145

\bibitem[{{Reiners} {et~al.}(2016){Reiners}, {Lemke}, {Bauer}, {Beeck}, \&
  {Huke}}]{Reiners2016}
{Reiners}, A., {Lemke}, U., {Bauer}, F., {Beeck}, B., \& {Huke}, P. 2016, A\&A,
  595, A26

\bibitem[{Ricchiazzi \& Canfield(1983)}]{Ricchiazzi1983ApJ272}
Ricchiazzi, P.~J. \& Canfield, R.~C. 1983, ApJ, 272, 739

\bibitem[{{Rimmele} {et~al.}(2020){Rimmele}, {Warner}, {Keil}, {Goode},
  {Kn{\"o}lker}, {Kuhn}, {Rosner}, {McMullin}, {Casini}, {Lin}, {W{\"o}ger},
  {von der L{\"u}he}, {Tritschler}, {Davey}, {de Wijn}, {Elmore}, {Fehlmann},
  {Harrington}, {Jaeggli}, {Rast}, {Schad}, {Schmidt}, {Mathioudakis},
  {Mickey}, {Anan}, {Beck}, {Marshall}, {Jeffers}, {Oschmann}, {Beard},
  {Berst}, {Cowan}, {Craig}, {Cross}, {Cummings}, {Donnelly}, {de Vanssay},
  {Eigenbrot}, {Ferayorni}, {Foster}, {Galapon}, {Gedrites}, {Gonzales},
  {Goodrich}, {Gregory}, {Guzman}, {Guzzo}, {Hegwer}, {Hubbard}, {Hubbard},
  {Johansson}, {Johnson}, {Liang}, {Liang}, {McQuillen}, {Mayer}, {Newman},
  {Onodera}, {Phelps}, {Puentes}, {Richards}, {Rimmele}, {Sekulic}, {Shimko},
  {Simison}, {Smith}, {Starman}, {Sueoka}, {Summers}, {Szabo}, {Szabo},
  {Wampler}, {Williams}, \& {White}}]{Rimmele2020}
{Rimmele}, T.~R., {Warner}, M., {Keil}, S.~L., {et~al.} 2020, SoPh, 295, 172

\bibitem[{{Sasso} {et~al.}(2011){Sasso}, {Lagg}, \& {Solanki}}]{Sasso2011}
{Sasso}, C., {Lagg}, A., \& {Solanki}, S.~K. 2011, A\&A, 526, A42

\bibitem[{{Scherrer} {et~al.}(2012){Scherrer}, {Schou}, {Bush}, {Kosovichev},
  {Bogart}, {Hoeksema}, {Liu}, {Duvall}, {Zhao}, {Title}, {Schrijver},
  {Tarbell}, \& {Tomczyk}}]{Scherrer2012}
{Scherrer}, P.~H., {Schou}, J., {Bush}, R.~I., {et~al.} 2012, SoPh, 275, 207

\bibitem[{{Schmidt} {et~al.}(2012){Schmidt}, {von der L{\"u}he}, {Volkmer},
  {Denker}, {Solanki}, {Balthasar}, {Bello Gonzalez}, {Berkefeld}, {Collados},
  {Fischer}, {Halbgewachs}, {Heidecke}, {Hofmann}, {Kneer}, {Lagg}, {Nicklas},
  {Popow}, {Puschmann}, {Schmidt}, {Sigwarth}, {Sobotka}, {Soltau}, {Staude},
  {Strassmeier}, \& {Waldmann }}]{schmidt2012a}
{Schmidt}, W., {von der L{\"u}he}, O., {Volkmer}, R., {et~al.} 2012, AN, 333,
  796

\bibitem[{{Sowmya} {et~al.}(2022){Sowmya}, {Lagg}, {Solanki}, \& {Castellanos
  Durán}}]{Sowmya2022}
{Sowmya}, K., {Lagg}, A., {Solanki}, S.~K., \& {Castellanos Durán}, J.~S.
  2022, A\&A, 661, A122

\bibitem[{Stix(2002)}]{2002tsai.book.....S}
Stix, M. 2002, The Sun, 2nd edn., Astronomy and Astrophysics Library (Springer
  Berlin, Heidelberg), XVI + 492, springer Book Archive

\bibitem[{{\v{S}}vestka(1976)}]{Svestka1976}
{\v{S}}vestka, Z.~F. 1976, Astrophysics and Space Science Library, Vol.~54,
  Solar Flares (Dordrecht: Springer)

\bibitem[{Tei {et~al.}(2018)Tei, Sakaue, Okamoto, Kawate, Heinzel, UeNo, Asai,
  Ichimoto, \& Shibata}]{tei2018}
Tei, A., Sakaue, T., Okamoto, T.~J., {et~al.} 2018, PASJ, 70, 100

\bibitem[{{Verma} {et~al.}(2012){Verma}, {Balthasar}, {Deng}, {Liu}, {Shimizu},
  {Wang}, \& {Denker}}]{Verma2012}
{Verma}, M., {Balthasar}, H., {Deng}, N., {et~al.} 2012, A\&A, 538, A109

\bibitem[{{Vicente Arévalo, Andrés} {et~al.}(2023){Vicente Arévalo,
  Andrés}, {Štěpán, Jiří}, {del Pino Alemán, Tanausú}, \& {Martínez
  González, María Jesús}}]{Vicente2023}
{Vicente Arévalo, Andrés}, {Štěpán, Jiří}, {del Pino Alemán, Tanausú},
  \& {Martínez González, María Jesús}. 2023, A\&A, 675, A45

\bibitem[{{W{\"o}ger} \& {von der L{\"u}he}(2008)}]{Woeger2008}
{W{\"o}ger}, F. \& {von der L{\"u}he}, O. 2008, in {Proc.\ SPIE}, Vol. 7019,
  {Advanced Software and Control for Astronomy II}, ed. A.~{Bridger} \& N.~M.
  {Radziwill}, 70191E

\bibitem[{{Xu} {et~al.}(2012){Xu}, {Lagg}, {Solanki}, \& Liu}]{Xu2012}
{Xu}, Z., {Lagg}, A., {Solanki}, S., \& Liu, Y. 2012, ApJ, 749, 138

\bibitem[{Zharkova {et~al.}(2011)Zharkova, Arzner, Benz, Browning, Dauphin,
  Emslie, Fletcher, Kontar, Mann, Onofri, Petrosian, Turkmani, Vilmer, \&
  Vlahos}]{Zharkova2011}
Zharkova, V.~V., Arzner, K., Benz, A.~O., {et~al.} 2011, Space Sci. Rev., 159,
  357

\end{thebibliography}

\end{document}